# Structure and magnetic properties of the S = 3/2 zigzag spin chain antiferromagnet BaCoTe$_2$O$_7$

Lisi Li,[1] Xunwu Hu,[1] Zengjia Liu,[1] Jia Yu,[1] Benyuan Cheng,[2,3] Sihao Deng,[4] Lunhua He,[4,5,6] Kun Cao,[1] Dao-Xin Yao,[1] and Meng Wang[1,*]

[1]Center for Neutron Science and Technology, School of Physics,
Sun Yat-Sen University, Guangzhou, 510275, China

[2]Shanghai Institute of Laser Plasma, Shanghai, 201800, China

[3]Center for High Pressure Science and Technology Advanced Research, Pudong,
Shanghai 201203, China

[4]Spallation Neutron Source Science Center, Dongguan, 523803, China

[5]Beijing National Laboratory for Condensed Matter Physics,
Institute of Physics, Chinese Academy of Sciences, Beijing 100190, China

[6]Songshan Lake Materials Laboratory, Dongguan 523808, China

We report an investigation on structure and magnetic properties of the S = 3/2 zigzag spin chain compound BaCoTe$_2$O$_7$. Neutron diffraction measurements reveal BaCoTe$_2$O$_7$ crystallizes in the noncentrosymmetric space group *Ama2* with a canted ↑↑↓↓ spin structure along the quasi-one-dimensional zigzag chain and a moment size of 1.89(2) $\mu_B$ at 2 K. Magnetic susceptibility and specific heat measurements yield an antiferromagnetic phase transition at $T_N$ = 6.2 K. A negative Curie-Weiss temperature $\Theta_{CW}$ = −74.7(2) K and an empirical frustration parameter of $f = |\Theta_{CW}|/T_N \approx 12$ is obtained from fitting the magnetic susceptibility, indicating antiferromagnetic interactions and strong magnetic frustration. By employing ultraviolet-visible absorption spectroscopy and first principles calculations, an indirect band gap of 2.68(2) eV is determined. We propose that the canted zigzag spin chain of BaCoTe$_2$O$_7$ may produce a change of the polarization *via* exchange striction mechanism.

## I. INTRODUCTION

Quasi-one-dimensional (1D) systems have been extensively studied due to their intriguing properties [1–5]. An ideal 1D system does not form magnetic order at finite temperature [6]. However, interchain interactions are inevitable in a real 1D material, which would result in a three-dimensional (3D) magnetic ordered ground state [5]. Moreover, magnetic anisotropy arising from the spin orbital coupling and crystal field environment are another degrees of freedom that influence the magnetic order of a 1D system. In the spin-3/2 Co$^{2+}$ based 1D system, rich phase diagrams and some emergent phenomena appear under magnetic field or pressure owing to their strong magnetic anisotropy. A complicated magnetic field induced phase diagram was observed in Co$^{2+}$ based Ising spin chain systems BaCo$_2$V$_2$O$_8$[7,8] and CoNb$_2$O$_6$[9,10]. A plateau was observed in magnetization of the spin chain systems α-CoV$_2$O$_6$[11,12] and Ca$_3$Co$_2$O$_6$[13,14]. A combination of intra- and inter-chain interactions and magnetic anisotropy could derive various magnetic properties for a Co$^{2+}$ based 1D system. It is thus interesting to investigate the magnetic behaviors of the Co$^{2+}$ based 1D systems for exploring exotic physics.

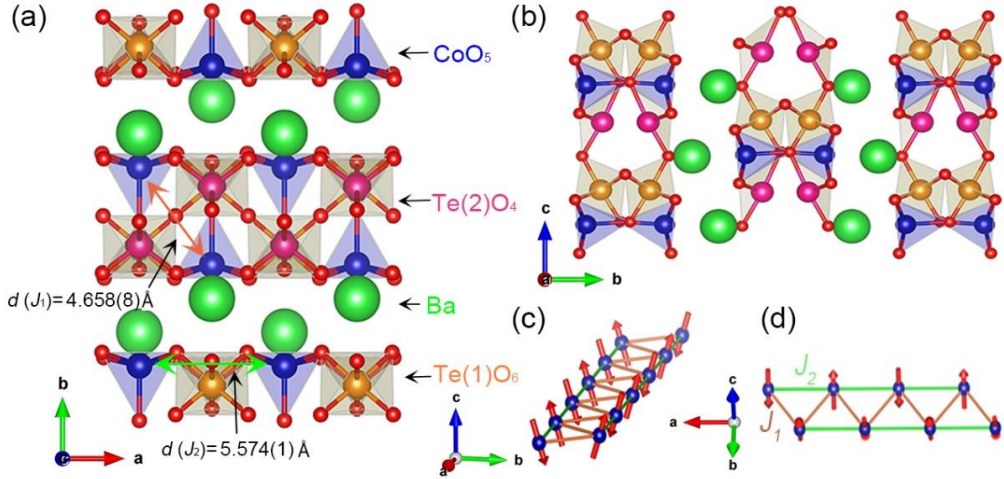

FIG. 1: The crystal structure of BaCoTe$_2$O$_7$ viewed along (a) the *c* and (b) the *a* direction. (c) and (d) Magnetic structure and exchange interactions network in a zigzag chain in different views. CoO$_5$, Te(1)O$_6$, and Te(2)O$_4$ and Ba atoms are drawn by different colors. The paths of the intrachain interactions $J_1$ and $J_2$ are marked by orange and green, respectively. The red arrows represent the spin directions.

Ba*M*Te$_2$O$_7$ (*M* = Cu, Zn, Mg) belong to a 1D chain system and crystallize in the noncentrosymmetric space group *Ama2* (No. 40)[15,16]. Their structures highlight a zigzag chain arrangement of the transition metal atoms *M* with the crystal field environment *M*O$_5$, in which *M*O$_5$ are not directly connected by O$^{2+}$ as shown in Fig. 1(a). Ba*M*Te$_2$O$_7$ (*M* = Zn, Mg) are nonmagnetic [16]. Although the spin of Cu$^{2+}$ is 1/2 in BaCuTe$_2$O$_7$, no magnetic order has been detected down to 1.8 K [15]. Similar Co$^{2+}$ based zigzag spin chain compounds α-BaCo*X*$_2$O$_7$ (*X* = P, As) have been extensively investigated [17,18]. They show antiferromagnetic (AFM) orders below $T_N$ =11 and 10 K, respectively. α-BaCo*X*$_2$O$_7$ exhibits large magnetic anisotropy on susceptibility. Their magnetization exhibits a 1/3 plateau below $T_N$, which is resulted from their incommensurate atomic shifts. Given the fascinating magnetic properties in the Co$^{2+}$ based zigzag spin chain system, it is interesting to study the properties of BaCoTe$_2$O$_7$. There are no reports on the synthesis and characterization of BaCoTe$_2$O$_7$ so far to our knowledge.

TABLE I: Atoms, Wyckoff site, fractional atomic coordinates ($x/a$, $y/b$, $z/c$) and occupation of BaCoTe$_2$O$_7$ at 2 K. The numbers in the brackets are the corresponding errors. The refined lattice constants are $a$ = 5.5685(1), $b$ = 15.1449(4), and $c$ = 7.2750(2) Å and profile parameters are $R_p$ = 4.83%, $R_{wp}$ = 4.68%, $R_e$ = 1.75%, and $\chi^2$ = 7.19.

| Atoms | Wyckoff site | x/a | y/b | z/c | Occupation |
| --- | --- | --- | --- | --- | --- |
| Ba | 4b | 1/4 | 0.2083(2) | 0.003(1) | 1.0 |
| Co | 4b | 1/4 | 0.1246(3) | 0.519(2) | 0.97(2) |
| Te1 | 4b | 1/4 | 0.9230(2) | 0.765(1) | 0.95(2) |
| Te2 | 4b | 3/4 | 0.0719(2) | 0.250(1) | 0.95(2) |
| O1 | 4b | 3/4 | 0.1349(1) | 0.006(1) | 0.97(1) |
| O2 | 8c | 0.0028(8) | 0.1438(1) | 0.3115(9) | 1.98(3) |
| O3 | 4b | 1/4 | 0.9856(2) | 0.5357(8) | 0.93(2) |
| O4 | 8c | 0.0065(8) | 0.8428(1) | 0.6935(9) | 1.85(3) |
| O5 | 4a | 0 | 0 | 0.877(1) | 0.91(2) |

In this present study, magnetic susceptibility, magnetization, specific heat, neutron powder diffraction (NPD), ultraviolet-visible (UV-vis) absorption spectroscopy measurements, and first principle density functional theory (DFT) calculations were performed to uncover the properties of BaCoTe$_2$O$_7$. A canted ↑↑↓↓ spin structure is identified below $T_N$ = 6.2 K with a propagation vector $\boldsymbol{k}$ = (0.5, 0, 0). A negative Curie-Weiss temperature $\Theta_{CW}$= −74.7(2) K obtained from fitting the susceptibility indicates that the AFM interactions dominate. Large anisotropy is revealed on susceptibility and magnetization measurements. An indirect band gap is determined as 2.68(2) eV. The ordered moment of Co$^{2+}$ may be reduced by covalency between the Co$^{2+}$ 3$d$ and O$^{2-}$ 2$p$ orbitals and magnetic frustration.

## II. EXPERIMENTS AND CALCULATIONS

Single crystalline samples of BaCoTe$_2$O$_7$ were grown by two steps. First, a polycrystalline sample was synthesized using the conventional solid-state reaction. The starting materials BaCO$_3$ (99.99%), Co$_3$O$_4$ (99.9%), and TeO$_2$ (99.99%) were mixed thoroughly in the stoichiometric ratio in an agate mortar. The mixed powders were then pressed to a pellet and put into an alumina crucible covered with a lid. The pellet was calcined at 650 °C in air in a muffler furnace for 7 days with several intimidate grindings. Second, singlecrystalline samples were grown using the flux method. NaCl and KCl mixed in a 1:1 molar ratio were used as flux. The BaCoTe$_2$O$_7$ powder samples and fluxes with a weight ratio 1:0.75 were placed to an alumina crucible. The mixtures were heated to 800 °C within 6 hours and dwelled for 24 hours.

After then, it was cooled to 650 °C at 1 °C/h then down to room temperature by tuning off the muffler furnace. The flux was washed with hot water. Single crystal x-ray diffraction (XRD) measurement was performed on a SuperNova (Rigaku) single-crystal x-ray diffractometer.

Our NPD experiment was conducted on the general purpose powder diffractometer (GPPD) installed at the China spallation neutron source (CSNS) [19]. The powder samples of BaCoTe$_2$O$_7$ were filled in a cylindrical vanadium can and measured at 2 and 300 K. The Rietveld method was employed to refine the NPD patterns using the *FullProf* suit package [20]. DC susceptibility and specific heat data were collected on a commercial physical property measurement system (PPMS, Quantum Design). The direction of the single crystal was determined by a x-ray Laue diffractometer. UV-vis spectroscopy measurements were conducted on an Ocean Optics DH-2000-BAL spectrometer.

We performed DFT calculations using the projector augmented-wave [21] method with generalized gradient approximation (GGA) of Perdew-Burke-Ernzerhof [22] functional as implemented in the Vienna *ab initio* simulation package [23,24]. The kinetic cutoff energy was set to 550 eV and Γ-centered Monkhorts-Pack meshes of 2×2×3 was used for Brillouin-zone integrations [25]. For comparison with experiment, the lattice constants and spin structure refined from NPD at 2 K were employed. The atomic positions were fully relaxed with a conjugated gradient algorithm until remaining force on each atom was less than 0.01 eV/angstrom. Spin-orbit coupling is taken into account in all calculations. We also calculated the free energy of the different magnetic configurations.

## III. RESULTS AND DISCUSSION

The observed and Calculated NPD patterns of BaCoTe$_2$O$_7$ are shown in Fig. 2. The refined structure can well reproduce the NPD patterns at 300 and 2 K with considering magnetic reflections. No impurity phase is found. Our refinement reveal BaCoTe$_2$O$_7$ crystallizes in space group *Ama2* (No. 40) and isostructurally to Ba*M*Te$_2$O$_7$ (*M* = Cu, Zn, Mg). The obtained atomic coordinates and some selected bond lengths at 2 K are tabulated in Table I. The composition from the NPD refinement is BaCo$_{0.97}$Te$_{1.9}$O$_{6.64}$, close to the stoichiometric content. The refined parameters at 300 K are presented in supplemental material. As shown in Fig. 1, the magnetic ions Co$^{2+}$ is surrounded by five O$^{2-}$ anions forming a CoO$_5$ square pyramid crystal field environment. Besides, Te atoms host two valence states Te(1)$^{6+}$ and Te(2)$^{4+}$, which involves the Te(1)O$_6$ and Te(2)O$_4$ crystal field environments, respectively. The CoO$_5$, Te(1)O$_6$, and Te(2)O$_4$ are lined up in a zigzag chain manner running along the *a* direction. Along the *b* direction, Ba$^{2+}$ ions separate the [CoTe$_2$O$_7$]$^{2-}$ two dimensional layers. For the *c* direction, three zigzag chains stack as CoO$_5$, Te(1)O$_6$, and Te(2)O$_4$ periodic sequence. The crystal structure is further confirmed by a single crystal XRD experiment (See supplemental material).

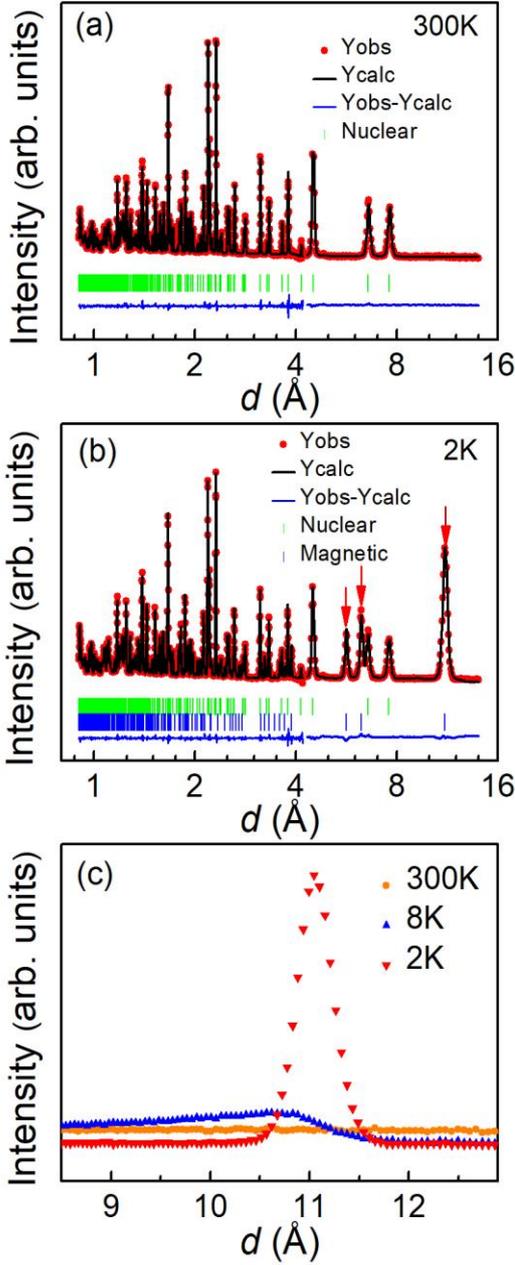

FIG. 2: Observed and calculated NPD patterns of BaCoTe$_2$O$_7$ at (a) 300 and (b) 2 K. (c) A zoom-in version of the NPD pattern in the range between 8 and 12.5 Å.

TABLE II: Basic vectors of magnetic Co site with the propagation vector $k$ = (0.5,0,0). The atom sites of Co(1) and Co(2) are ($x, y, z$) and ($-x + 1, -y + 1/2, z - 1/2$), respectively.

| IR | BV | Co (1) | Co (2) |
|---|---|---|---|
| Γ1 | $\psi_1$ | [1,0,0] | [-1,0,0] |
|    | $\psi_2$ | [0,1,0] | [0,-1,0] |
|    | $\psi_3$ | [0,0,1] | [0,0,1] |

Reflections from magnetic scattering could be identified clearly in Fig. 2 (b) by comparing the NPD patterns at 300 and 2 K. All of these magnetic peaks can be indexed with a propagation vector $k$ = (0.5, 0, 0). A representational analysis was performed to quantitatively determine the spin structure using the BasIreps program in the *FullProf* suite package [26,27], which results in a nonzero irreducible representation (IR) Γ1. One crystallographically site of Co$^{2+}$ ion is splitted into two sites Co$^{2+}$ (1) ($x, y, z$) and Co$^{2+}$ (2) ($-x+1, -y+1/2, z-1/2$). Basis vectors (BV) for the two Co$^{2+}$ sites of the IR Γ1 are displayed in Table II. The refinement with IR Γ1 can well reproduce the magnetic peaks at 2 K with the magnetic $R$ factor $R_{mag}$ = 4.27%. As a result, the moment components along the $a$, $b$, and $c$ directions are $m_a$= 0(0.1), $m_b$= 0.60(4), and $m_c$= 1.79(2) $\mu_B$, yielding a total magnetic moment $m_{tot}$ = 1.89(2) $\mu_B$. The determined spin structures in different views are drawn in Figs. 1 (c) and (d). The spin is arranged within the $bc$ plane perpendicular to the zigzag chain direction. Along the zigzag chain, the adjacent spins exhibit an angle of 37° and is arrayed in an alternative ferromagnetic (FM) and AFM way, which can also be recognized as a canted ↑↑↓↓ spin structure. A similar magnetic structure with only the moment component along the $c$ axis is also tested, yield $m$ = 1.86(1) $\mu_B$ and $R_{mag}$= 4.69%, which is worse than the canted spin structure. Figure 2 (c) displays NPD patterns for a magnetic peak at 2, 8, and 300 K. The broad peak at 8 K that is above the Néel temperature suggest existence of the magnetic frustration and short range magnetic correlation.

Figure 3(a) shows temperature dependence of magnetic susceptibility $\chi$ measured under zero field cooling (ZFC) and field cooling (FC) on a powder sample of BaCoTe$_2$O$_7$. No obvious difference is found between the ZFC and FC data. There is a broad peak around $T_{max} \approx 20$ K, which is a commonly observed characteristic for a 1D magnet due to the presence of magnetic correlation [5]. In the derivative $d\chi/dT$ in the inset of Fig. 3(a), a peak appears at $T_N$ = 6.2 K, which could be attributed to the AFM phase transition. A fitting of the inverse susceptibility $1/(\chi - \chi_0)$ in temperature range from 100 to 300 K to the Curie-Weiss law $\chi = \chi_0 + C/(T - \Theta_{CW})$ is depicted as the red solid line in Fig. 3(a). The fitting yields that the contributions of core diamagnetism and Van Vleck paramagnetism $\chi_0$ amounts to $-7.63(8) \times 10^{-4}$ emu Oe$^{-1}$ mol$^{-1}$, the Curie constant $C$ is 4.22(1) emu K mol$^{-1}$ Oe$^{-1}$, and the Curie-Weiss temperature $\Theta_{CW}$ is $-74.7(2)$ K. The negative $\Theta_{CW}$ demonstrates that AFM interactions dominate in the system. An empirical formula of the frustrated parameter is $f = |\Theta_{CW}|/T_N = $ 12.08 > 10, suggesting the possible existence of magnetic frustration or effect of low dimensionality [28]. The effective magnetic moment is estimated to be $\mu_{eff} = 5.77(1)\ \mu_B$, which is larger than that of the theoretical value of $\mu_{eff} = g\sqrt{S(S+1)} = 3.87\ \mu_B$ for a S = 3/2 system when assuming zero orbital residual with $g = 2$. The results signal a considerable orbital contribution from the $Co^{2+}$ ions in consistence with the reports in other $Co^{2+}$ based materials, such as $BaCoX_2O_7$ (X = P, As) [18], $Na_3Co_2SbO_6$ [29], $BaCo_2(AsO_4)_2$ [30].

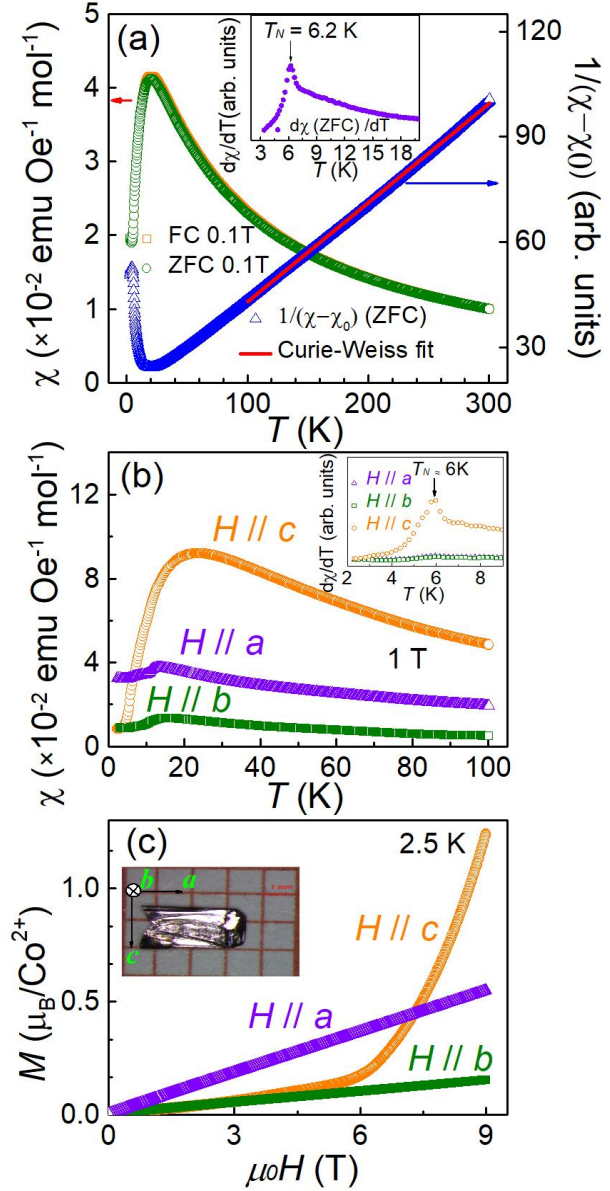

FIG. 3: (a) Left: Temperature dependence of susceptibility $\chi$ measured on a powder sample of BaCoTe$_2$O$_7$ and inset its derivative $d\chi/dT$, Right: Inverse susceptibility $1/(\chi - \chi_0)$ and a Curie-Weiss fit. (b) Temperature dependence of $\chi$ measured in three crystallographic directions of a single crystal and inset their derivative $d\chi/dT$. (c) Magnetization measured in three crystallographic directions of a single crystal. Inset shows a single crystal picture and the crystallographic directions.

Figures 3 (b) and (c) present the temperature dependence of magnetic susceptibility $\chi$ and field dependent magnetization $M$ measured in different crystallographic directions on a single crystal of $BaCoTe_2O_7$. The susceptibility measured with the magnetic filed applied along the $a$ direction $\chi_a$, the $b$ direction $\chi_b$, and the $c$ direction $\chi_c$ exhibit a phase transition at $T_N \approx 6$ K, as seen in the derivative $d\chi/dT$, in agreement with that measured on the powder sample. In addition, these data exhibit a large difference in the whole temperature range. For an isotropy 3D antiferromagnet, $\chi_a$, $\chi_b$, and $\chi_c$ should coincide in the high temperature range ($T > T_N$) [31]. The large anisotropy between $\chi_a$, $\chi_b$, and $\chi_c$ may be due to anisotropy of the spin correlations persisting above $T_N$ and the 1D nature of $BaCoTe_2O_7$. For example, the $Co^{2+}$ based 1D compounds $BaCoP_2O_7$ [17], $Ba_2CoSi_2O_7$ [32], and $BaCo_2V_2O_8$ [7] also present a large anisotropy. The kinks on $\chi_a$ and $\chi_b$ at ~13 K in Fig. 3 (b) may correspond to impurities in the sample contaminated during the single crystal growth process. The magnetization $M_c$ measured with a magnetic field applied along the $c$ direction exhibits a sharp change around $\mu_0 H_c = 6$ T while it shows no anomaly in $M_a$ and $M_b$. This is consistent with a spin flop-like transition for spins aligned in the $bc$ plane with a dominant moment component $m_c$.

We present temperature dependence of the specific heat $C_p$ measured under zero and magnetic fields on a pellet pressed from a powder sample of $BaCoTe_2O_7$ in Fig. 4. A λ-like phase transition is observed at $T_N = 6.2$ K, consistent with the susceptibility measurements in Fig. 3. To estimate magnetic contribution $C_{p,mag}$ on the specific heat, the phonon part $C_{p,ph}$ is subtracted from the total $C_p$, while electronic contribution is not considered since $BaCoTe_2O_7$ is an insulator. A modified Debye model is taken into account to estimate the phonon contribution considering two phonon spectra including the spectrum of the light atoms ($O^{2-}$) and the heavy atoms ($Ba^{2+}$, $Co^{2+}$, $Te^{4+}$, and $Te^{6+}$). The method was proved effectively in $M_2MnTeO_6$ ($M$ = Sr, Ba) [33,34]. The formula of the modified Debye model is given by:

$$C_{p,ph} = 9R \sum_{n=1}^{2} C_n \left(\frac{T}{\Theta_{Dn}}\right)^3 \int_0^{\Theta_{Dn}/T} \frac{x^4 e^x}{(e^x - 1)^2} dx \quad (1)$$

A fitting of the raw data in temperature range from 60 to 200 K to Eq. (1) yields that 6.5 light atoms and 4.5 heavy atoms out of the 11 atoms in a formula unit of $BaCoTe_2O_7$. The Debye temperatures corresponding to the light and heavy atoms are $\Theta_{D1} = 855(26)$ K and $\Theta_{D2} = 256(7)$ K, respectively. The ratio is in good agreement with 7 light atoms ($O^{2-}$) and 4 heavy atoms ($Ba^{2+}$, $Co^{2+}$, $Te^{4+}$, and $Te^{6+}$) in the formula unit of $BaCoTe_2O_7$. Based on the fitted parameters, the phonon part specific heat $C_{p,ph}$ is extrapolated to 1.8 K as the black dashed line plotted in

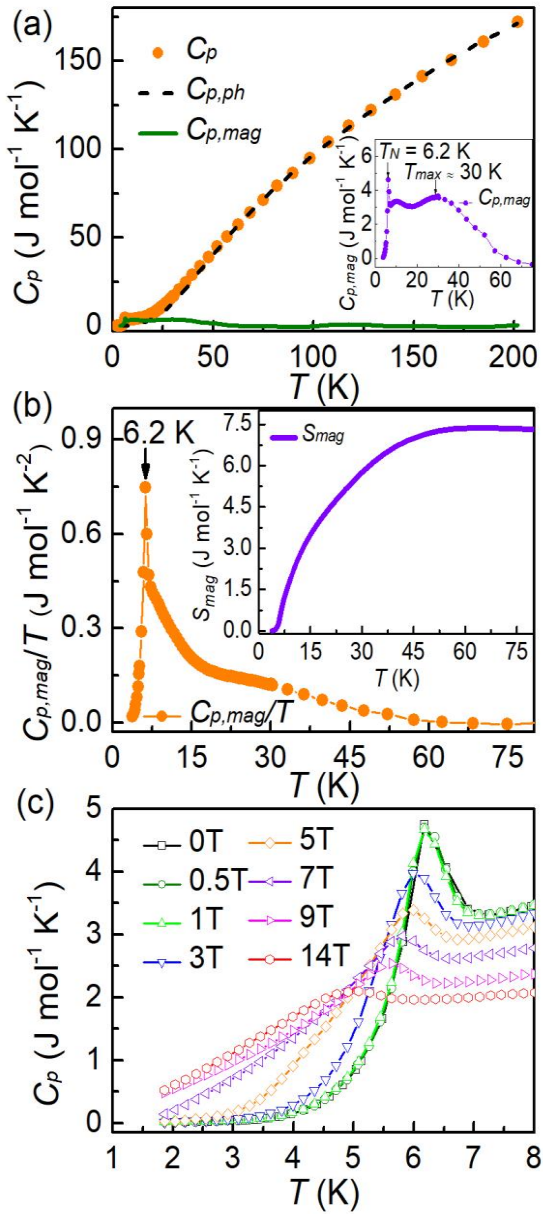

FIG. 4: (a) Temperature dependence of the total $C_p$, phonon part $C_{p,ph}$, and magnetic part $C_{p,mag}$ of the specific heat of BaCoTe$_2$O$_7$. The inset shows the zoom-in of $C_{p,mag}$ at low temperature. (b) Derived $C_{p,mag}/T$ against temperature with an inset presenting the magnetic entropy. (c) Specific heat $C_p$ measured with different applied magnetic fields in temperature range of 1.8 − 8 K.

Fig. 4(a). The magnetic part of specific heat $C_{p,mag}$ is obtained accordingly and depicted as the green solid line. A broad peak with a maximum at ∼ 30 K on $C_{p,mag}$ in the inset figure of Fig. 4 (a) reveals the persistence of the magnetic correlation above $T_N$, consistent with the magnetic susceptibility measurements in Fig. 3. The magnetic entropy contribution to the magnetic state changes can be deduced from the formula $S_{mag}=\int C_{p,mag}/TdT$. As shown in Fig. 4 (b), the magnitude of the obtained magnetic entropy $S_{mag}$ = 7.38 J mol$^{-1}$ K$^{-1}$ at 80 K as is lower than the theoretical formula of $S_{mag}$ = R ln (2$S$ + 1) = 11.52 J mol$^{-1}$ K$^{-1}$ for a magnetic correlation free Co$^{2+}$ ion in the spin only S = 3/2 case. The reduced $S_{mag}$ could be ascribed to the reduction of the ordered moment and the existence of the magnetic correlations well above $T_N$. Figure 4 (c) presents magnetic field dependence of the Néel temperature, where $T_N$ shifts toward to lower temperatures with increasing magnetic fields. The similar behavior has also been observed in the zigzag chain antiferromagnet BaNd$_2$O$_4$ [35].

To get deeper insight on the experimental results, we performed DFT calculations to study the electronic structure and magnetism of BaCoTe$_2$O$_7$. The band gap is 2.68(2) eV determined from the UV-vis absorption spectrum. A Coulomb repulsive U was chosen as 6 eV accordingly. Figure 5 presents the calculated density of states (DOS) and electronic band structure of BaCoTe$_2$O$_7$. In Fig. 5 (a), it is shown that both the partial DOS of Co$^{2+}$ and O$^{2-}$ have significant contributions in the energy range of $-5 \leq E \leq -1$ eV, indicating a covalency between the 3$d$ and 2$p$ orbitals of Co$^{2+}$ and O$^{2-}$, respectively. The energy for the AFM ↑↑↓↓ spin structure -5.8146 eV/atom is lower than that of the FM spin structure -5.7801 eV/atom. The calculated ordered moment is $m_a = 0$, $m_b = 0.89$, $m_c = 2.66$, and $m_{tot} = 2.80$ $\mu_B$ by assuming the experimentally determined spin structure. The calculated ordered moment is smaller than the ideal local moment of 3 $\mu_B$ for Co$^{2+}$ with a 3$d^7$ electronic configuration, which may be caused by the covalency. But it is much larger than the experimental observation. The Co$^{2+}$ ions in BaCoTe$_2$O$_7$ arrange in a triangular lattice in the zigzag chain and the competition between intrachain exchange interactions $J_1$ ($d = 4.658(8)$ Å at $T = 2$ K) and $J_2$ ($d = 5.574(1)$ Å at $T = 2$ K) may lead to strong magnetic frustration, resulting in the reduced moment 1.89(2) $\mu_B$ as observed in experiment. We note weak interchain exchange interactions $J_3$ are necessary to stabilize the long range magnetic order.

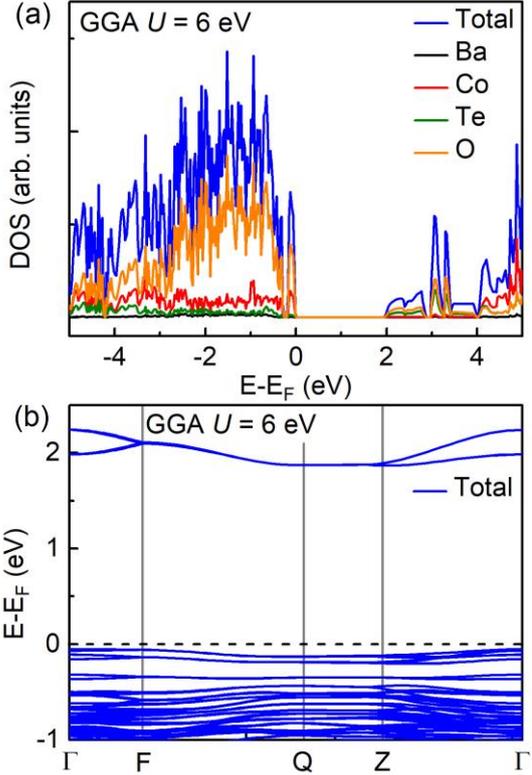

FIG. 5: (a) Calculated total and partial DOS for BaCoTe$_2$O$_7$ and (b) the band structure with the experimental determined magnetic structure and U = 6 eV.

The noncentrosymmetric BaCoTe$_2$O$_7$ is pyroelectric in its paramagnetic phase, but not ferroelectric, since its polarization can not be reversed by an applied electric field [15,16]. However, in the magnetically ordered phase, the spin structure with alternative FM and AFM arrangements may further lower the symmetry and result in a change of the electric polarization ΔP along the chain direction, by inducing different exchange striction on the ↑↑ and ↑↓ spin pairs. ΔP may be partially reversible, which then presents a scenario similar to type II multiferroics [36–41]. Therefore, it is worth-while to explore magnetoelectric coupling in BaCoTe$_2$O$_7$, which we will leave for future work.

## IV. CONCLUSIONS

In summary, we have synthesized both powder and single crystal samples of BaCoTe$_2$O$_7$ and investigated the structure, magnetic and electronic properties. An antiferromagnetic ↑↑↓↓ spin structure with a propagation vector $\mathbf{k} = (0.5, 0, 0)$ is revealed below $T_N = 6.2$ K. The large magnetic anisotropy at high temperatures suggests the existence of the magnetic exchange interactions above $T_N$. These field-induced behaviors could be attributed to the effects of the single ion anisotropy, magnetic exchange interactions of $Co^{2+}$, and the 1D nature of this compound. Our DFT calculations suggest the existence of covalency between the $Co^{2+}$ $3d$ and $O^{2-}$ $2p$ orbits. The covalency and magnetic frustration in the system may reduce its ordered moment. Further investigations on spin dynamics and multiferroics in $BaCoTe_2O_7$ are meaningful for reveling its unique magnetic and electric properties.

## V. ACKNOWLEDGEMENTS

We appreciate Hanjie Guo for the Laue measurement. M. W. was supported by the National Natural Science Foundation of China (Grant No. 11904414), National Key Research and Development Program of China (No. 2019YFA0705702). D. X. Y. was supported by NKRDPC-2018YFA0306001, NKRDPC-2017YFA0206203, NSFC-11974432, GBABRF-2019A1515011337, and Leading Talent Program of Guangdong Special Projects.

# Supplemental material: Structure and magnetic properties of the S=3/2 zigzag spin chain antiferromagnet BaCoTe$_2$O$_7$

Lisi Li[1], Xunwu Hu[1], Zengjia Liu[1], Jia Yu[1], Benyuan Cheng[2,3], Sihao Deng[4], Lunhua He[4,5,6], Kun Cao[1], Dao-Xin Yao[1], and Meng Wang[1, *]

[1]Center for Neutron Science and Technology, School of Physics,
Sun Yat-Sen University, Guangzhou, 510275, China

[2]Shanghai Institute of Laser Plasma, Shanghai, 201800, China

[3]Center for High Pressure Science and Technology Advanced Research, Pudong, Shanghai 201203, China

[4]Spallation Neutron Source Science Center, Dongguan, 523803, China

[5]Beijing National Laboratory for Condensed Matter Physics,
Institute of Physics, Chinese Academy of Sciences, Beijing 100190, China

[6]Songshan Lake Materials Laboratory, Dongguan 523808, China

## S1. Single crystal diffraction results

The single crystal diffraction was performed on a SuperNova (Rigaku) single-crystal x-ray diffractometer using a crystal with size 0.03×0.04×0.06 mm$^3$. The crystal structure was solved by the ShelXT-97 program in the Olex2 software. Table S1 presents the single crystal refinement and Table S2 shows the atomic coordinates and isotropic displacement parameters.

Table S1. Single crystal refinement of BaCoTe$_2$O$_7$.

| Empirical formula | Ba$_4$Co$_4$O$_{28}$Te$_8$ |
|---|---|
| Formula weight | 2253.88 |
| Temperature/K | 150.0(1) |
| Crystal system | orthorhombic |
| Space group | Ama2 (No.40) |

| a, b, c /Å | 5.5722(2), 15.1470(4), 7.2756(2) |
| --- | --- |
| $\alpha, \beta, \gamma$/° | 90, 90, 90 |
| Volume/Å$^3$ | 614.08(3) |
| $\rho_{calc}$ g/cm$^3$ | 6.095 |
| F (000) | 972.0 |
| Crystal size/mm$^3$ | 0.03 × 0.04 × 0.06 |
| 2Θ range for data collection/° | 9.082 to 60.41 |
| Index ranges | -7 ≤ h ≤ 7, -20 ≤ k ≤ 21, -9 ≤ l ≤ 9 |
| Reflections collected | 4896 |
| Independent reflections | 941 [R$_{int}$ = 0.0354, R$_{sigma}$ = 0.0247] |
| Data/restraints/parameters | 941/1/61 |
| Goodness-of-fit on F2 | 1.096 |
| Final R indexes [I>=2σ (I)] | R$_1$ = 1.60%, w$_{R2}$ = 3.73% |
| Final R indexes [all data] | R$_1$ = 1.63%, w$_{R2}$ = 3.75% |
| Largest diff. peak/hole / eÅ$^{-3}$ | 1.05/-1.27 |

Table S2. Fractional atomic coordinates and equivalent isotropic displacement parameters for BaCoTe$_2$O$_7$ at 150.0(1) K. $U_{iso}$ is defined as 1/3 of the trace of the orthogonalized $U_{IJ}$ tensor. The lattice constants are a=5.5722(2), b=15.1470(4), and c=7.2756(2) Å. The refined profile factors are R$_1$ =1.63%, w$_{R2}$ = 3.75%.

| Atoms | x/a | y/b | z/c | $U_{iso}$ (×10$^{-3}$Å$^2$) |
| --- | --- | --- | --- | --- |
| Ba | 3/4 | 0.20792(3) | 0.9093(1) | 5.2(1) |
| Co | 1/4 | 0.37674(7) | 0.8945(3) | 4.6(2) |
| Te1 | 3/4 | 0.42835(4) | 0.16354(6) | 4.0(2) |
| Te2 | 3/4 | 0.42266(4) | 0.64790(6) | 3.3(2) |
| O1 | 3/4 | 0.3650(3) | 0.406(2) | 8.0(1) |
| O2 | 0.496(1) | 0.3562(3) | 0.0989(6) | 8.0(1) |
| O3 | 3/4 | 0.4850(4) | 0.8755(9) | 5.0(1) |
| O4 | 0.508(1) | 0.3434(3) | 0.7167(6) | 5.0(1) |
| O5 | 1 | 0.5 | 0.5352(8) | 6.0(1) |

## S2. Neutron powder diffraction pattern refinement

Table S3 presents the fractional atomic coordinates and isotropic displacement parameters at 300 K obtained from neutron powder diffraction (NPD) refinement. The results of 300K NPD refinement are shown in Fig. 2(a) in the text. The fractional atomic coordinates obtaining from NPD refinement are different from those obtaining from single crystal refinement which is basically due to the different selected coordinate axis of the two refinement.

Fig. S1 presents the magnetic structure refinement of the 2K NPD pattern using two models. From the refinement, we obtain that $m_c$ is the dominant component. In model 1, $m_a$ and $m_b$ are fixed to be 0 $\mu_B$, the magnetic peaks can be fulfilled with $m_c$=1.86(1) $\mu_B$, the profiled factors are $R_p$=4.91%, $R_{wp}$=4.77%, $\chi^2 = 7.43$, $R_{mag}$=4.69%. In model 2, $m_a$ and $m_b$ are free parameters, the refinement gives that $m_a$=-0.03(11), $m_b$=0.60(4), and $m_c$= 1.79(2) $\mu_B$, the profiled factors are $R_p$=4.83%, $R_{wp}$=4.68%, $\chi^2 = 7.19$, $R_{mag}$=4.27%. Model 2 is a better fit based on the refinement, we thus adapted the model 2 as the final results.

Table S3. Fractional atomic coordinates and isotropic displacement parameters at 300 K of BaCoTe$_2$O$_7$ obtain from NPD refinement. The lattice constants are $a$=5.5744(1), $b$=15.1979(4), $c$=7.2897(2) Å. The refined profile factors are $R_p$= 5.42%, $R_{wp}$= 4.81%, $R_e$=2.76%, and $\chi^2$= 3.042.

| Atoms | x/a | y/b | z/c | $U_{iso}$ (×10$^{-3}$Å$^2$) |
|---|---|---|---|---|
| Ba | 1/4 | 0.2082(2) | 0.004(1) | 1.0(8) |
| Co | 1/4 | 0.1228(3) | 0.518(2) | 7(2) |
| Te1 | 1/4 | 0.9231(3) | 0.7648(4) | 2(1) |
| Te2 | 3/4 | 0.0718(2) | 1/4 | 3(1) |
| O1 | 3/4 | 0.1346(1) | 0.006(1) | 3.3(7) |
| O2 | 0.0042(8) | 0.1435(1) | 0.3118(6) | 6.1(8) |
| O3 | 1/4 | 0.9853(2) | 0.5369(7) | 4(1) |
| O4 | 0.0074(7) | 0.8434(1) | 0.6941(6) | 2.9(7) |
| O5 | 0 | 0 | 0.8764(6) | 7.0(8) |

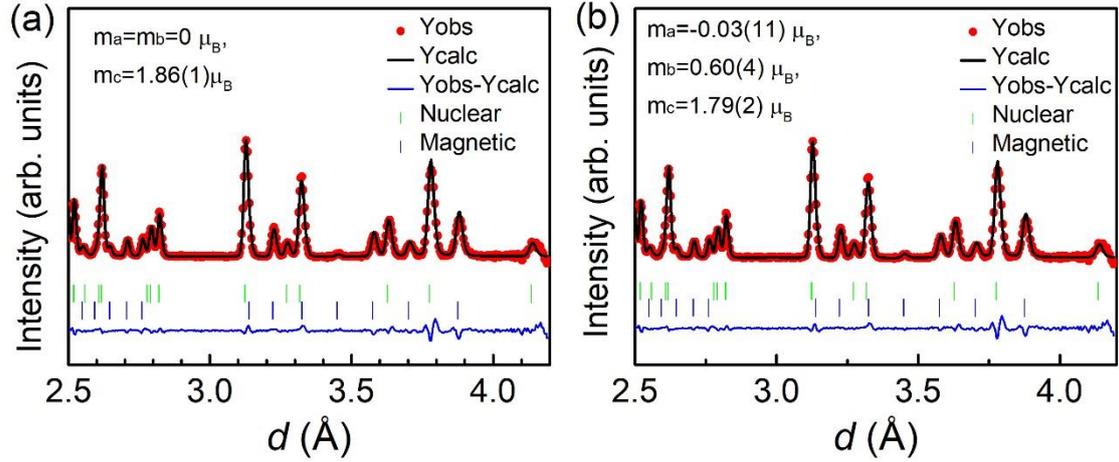

Fig. S1 Magnetic structure refinement of 2K NPD pattern with (a) $m_a$ and $m_b$ set to be 0 $\mu_B$, $m_c$ is a free parameter. The refined profiled factors are $R_p$=4.91%, $R_{wp}$=4.77%, $\chi^2 = 7.43$, $R_{mag}$=4.69%. (b) $m_a$, $m_b$, and $m_c$ are free parameters. The refined profiled factors are $R_p$=4.83%, $R_{wp}$=4.68%, $\chi^2 = 7.19$, $R_{mag}$=4.27%.

## S3. Energy dispersive x-ray spectra results

The energy dispersive x-ray (EDX) spectra were measured on an EDX spectroscopy (EVO, Zeiss). Two crystals were measured and five points were selected for collecting data for each crystal. The average atomic ratio for crystal 1 and 2 is Ba:Co:Te=1.00(1):1.10(1):2.04(1) and 1.00(1):1.10(2):2.03(1), respectively, close to the stoichiometric ratio. Fig. S2 presents a single crystal picture taking from the scanning electron microscope (SEM). Fig. S3 is a selected EDX spectrum. Table. S4 displays the atomic ratio of all results.

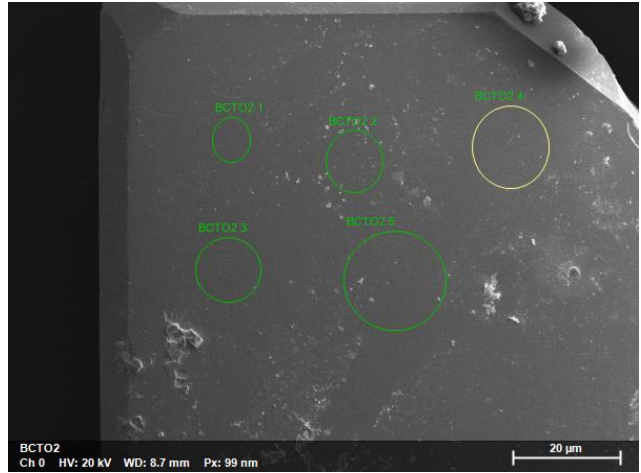

Fig. S2. Single crystal picture for BaCoTe$_2$O$_7$ takes from SEM.

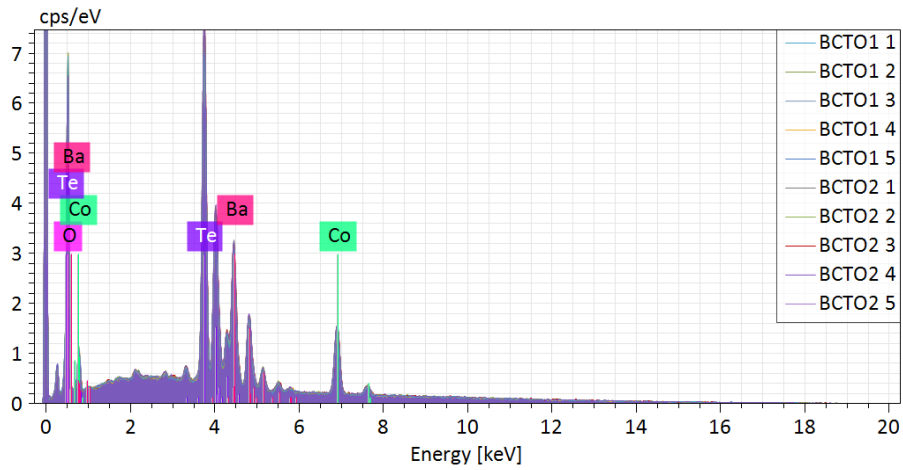

Fig. S3. An EDX spectrum of BaCoTe$_2$O$_7$ crystal.

Table S4. Atomic ratio obtains from EDX spectra for crystal (a) 1 and (b) 2 for BaCoTe$_2$O$_7$. The average atomic ratio for crystal 1 and 2 is Ba:Co:Te=1.00(1):1.10(1):2.04(1) and 1.00(1):1.10(2):2.03(1), respectively, close to the stoichiometric ratio of BaCoTe$_2$O$_7$.

(a) Atomic concentration [%]

| Spectrum | O | Co | Te | Ba |
|---|---|---|---|---|
| BCTO1 1 | 55.622 | 12.393 | 21.585 | 10.400 |
| BCTO1 2 | 55.868 | 11.760 | 21.599 | 10.773 |
| BCTO1 3 | 55.423 | 11.489 | 22.266 | 10.822 |
| BCTO1 4 | 55.260 | 11.798 | 22.181 | 10.761 |
| BCTO1 5 | 54.531 | 12.052 | 22.303 | 11.114 |
| Mean | 55.341 | 11.899 | 21.987 | 10.774 |
| Sigma | 0.506 | 0.341 | 0.363 | 0.254 |
| SigmaMean | 0.226 | 0.153 | 0.162 | 0.114 |

(b) Atomic concentration [%]

| Spectrum | O | Co | Te | Ba |
|---|---|---|---|---|
| BCTO2 1 | 52.778 | 12.697 | 23.227 | 11.298 |
| BCTO2 2 | 52.634 | 13.249 | 22.954 | 11.164 |
| BCTO2 3 | 52.805 | 11.946 | 23.625 | 11.624 |
| BCTO2 4 | 52.603 | 12.340 | 23.385 | 11.673 |
| BCTO2 5 | 52.646 | 12.748 | 23.230 | 11.375 |
| Mean | 52.693 | 12.596 | 23.284 | 11.427 |
| Sigma | 0.092 | 0.487 | 0.246 | 0.217 |
| SigmaMean | 0.041 | 0.218 | 0.110 | 0.097 |

## S4. UV-vis absorption spectroscopy measurements

The band gap of BaCoTe$_2$O$_7$ was measured using the UV-vis absorption spectroscopy. The band gap values was obtained by fitting the absorption spectra using the Tauc relation [1], i.e., $[\alpha \cdot h\nu]^{1/n} = A \cdot (h\nu - E_g)$, where $\alpha$ is the Kubelka–Mubelka–Munk function, $h$ is the Planck's constant, $\nu$ is the frequency, A is a constant, and $E_g$ is the band gap energy. The direct band gap is determined to be 2.98(2) eV with n = 1/2. The indirect band gap is determined to be 2.68(2) eV with n = 2. The fitting results are shown in Fig. S4. Based on the DFT calculations, BaCoTe2O7 is an indirect insulator.

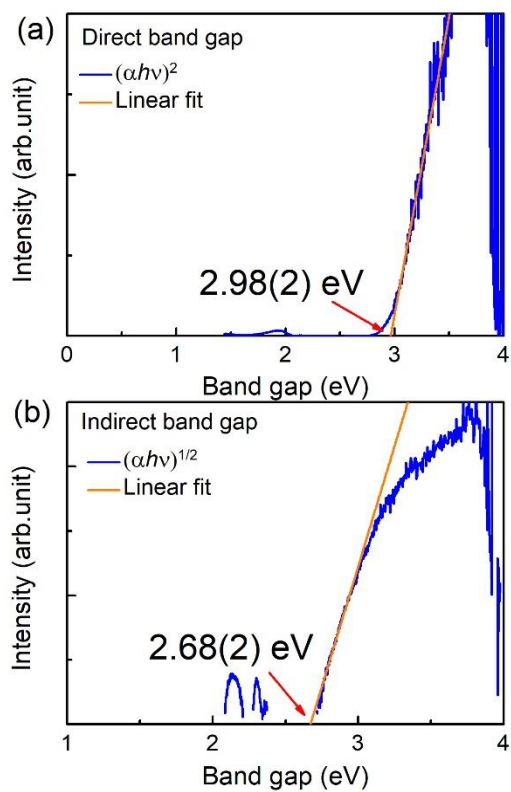

Fig. S4. UV-vis absorption spectroscopy measurements: (a) direct and (b) indirect band gap determination. The determined direct and indirect band gap are 2.98(2) and 2.68(2) eV, respectively.

[1] Sarkar, A. et al. Multicomponent equiatomic rare earth oxides with a narrow band gap and associated praseodymium multivalency. Dalton Trans. 46, 12167–12176 (2017).